\documentclass{cs20proc}

\editors{S. J. Wolk}
\publisher{Zenodo}
\conference{The 20th Cambridge Workshop on Cool Stars, Stellar Systems, and the Sun}
\conferencedate{2018}

\title{The Dynamics of OB Associations}
\author{Nicholas J. Wright$^{1}$}

\affiliation{$^{1}$Astrophysics Group, Keele University, Keele, ST5 5BG, UK}

\shorttitle{Dynamics of OB Associations}
\shortauthors{Nicholas J. Wright}

\abs{The formation and evolution of young star clusters and OB associations is fundamental to our understanding of the star formation process, the conditions faced by young binary and planetary systems, and the formation of long-lived open and globular clusters. Despite this our understanding of the physical processes that drive this evolution has been limited by the static nature of most observations. This is all changing thanks to a revolution in kinematic data quality from large-scale radial velocity surveys and new astrometric facilities such as {\it Gaia}. Here I summarise recent studies of multiple OB associations from both {\it Gaia} and ground-based astrometric surveys. These observations show that OB associations have considerable kinematic substructure and no evidence for the radial expansion pattern predicted by theories such as residual gas expulsion. This means that, contrary to the standard view of OB associations as expanded star clusters, these systems could never have been dense star clusters in the past and were most likely born as extended and highly substructured groups of stars. This places strong constraints on the primordial clustering of young stars and the conditions faced by young planetary systems.}

\begin{document}

\maketitle

\section{Introduction}

OB associations are low-density, co-moving groups of young stars that have been known about for over half a century \citep[e.g.,][]{amba47}. Though they were first identified because of their bright OB-type members, recent studies have shown them to have a fully-sampled mass function \citep[e.g.,][]{bric07,arms18}.

The low density of OB associations means they must be gravitationally unbound and therefore are likely to expand in the future \citep{amba49,blaa64}. This lead various authors to suggest that they might have undergone some expansion in the past and therefore formed as more compact structures \citep[e.g.,][]{blaa52,brow99}. In this scenario OB associations would have formed as compact star clusters, embedded within molecular clouds \citep{lada03}, that were then disrupted by some process such as residual gas expulsion \citep{hill80}. The disrupted remnant of the star cluster would then have dispersed into the Galactic field, being briefly visible as an unbound OB association \citep{lada03}.

In this contribution I will discuss recent studies that have attempted to test this model for the origins of OB associations. I will start by presenting both structural and kinematic studies of the most massive OB association known in our Galaxy, Cygnus OB2, and then present a kinematic study of the nearest OB association to the Sun, Scorpius-Centaurus.

\section{The Cygnus OB2 association}

Cyg OB2 is a massive OB association with a total stellar mass of $\sim$ (2--4) $\times 10^4$ M$_\odot$ \citep{drew08,wrig10a}. It is home to approximately 65 O-type stars, all spectroscopically studied and with masses up to $\sim$100~M$_\odot$ \citep{wrig15a}, providing a fully-sampled mass function. Its age is thought to be approximately 5~Myr, albeit with evidence for a considerable spread, as diagnosed from both the low-, intermediate- and high-mass stellar populations \citep{drew08,wrig10a,wrig15a}. The studies presented here are based on an X-ray selected sample of low-mass members \citep{wrig09a,wrig10a} and a recent spectroscopic census of high-mass members \citep{wrig15a}.

\subsection{The structure of Cyg OB2}

We used multiple quantitative diagnostics to study the structure of Cyg OB2 and search for evidence of mass segregation. We used the $Q$ parameter \citep{cart04}, a measure based on the minimum spanning tree, to search for evidence of substructure within the association. The $Q$ parameter varies from 0 -- 2, with low values indicating a highly substructured spatial distribution, while high values indicate a more centrally-condensed `cluster-like' distribution. To measure mass segregation we used both the mass segregation ratio, $\lambda_{\mathrm{MSR}}$ \citep{alli09}, and the local surface density ratio $\Sigma_{\mathrm{LDR}}$ \citep{masc11}. Both diagnostics provided similar results, so here we focus on those obtained from $\Sigma_\mathrm{LDR}$, for which values of $\sim 1$ indicate no mass segregation (i.e., the massive stars are located in regions of similar density to the low-mass stars) and larger values indicate mass segregation (the massive stars are located in regions of higher density to the low-mass stars, i.e., in the centres of dense groups or clusters).

\begin{figure}
	\centering
	\includegraphics[width=250pt]{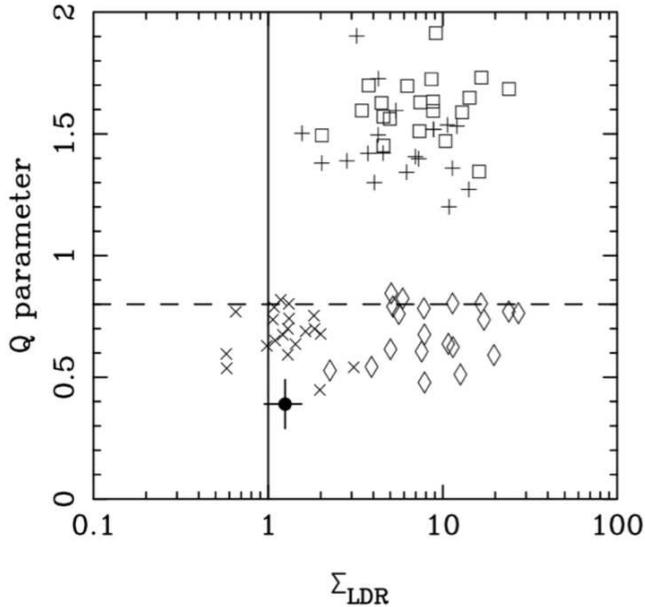}
	\caption{The $Q$ parameter plotted against $\Sigma_{\mathrm{LDR}}$ for Cygnus OB2 (black point) and for the results of N-body simulations at an age of 5~Myr that started from sub-virial (squares), virial (plus symbols), super-virial (diamonds) and low-density, super-virial (crosses) initial conditions. Figure adapted from results in \citet{wrig14b} and simulations from \citet{park14}.}
	\label{parameters}
\end{figure}

We measure $Q = 0.34 \pm 0.1$, which indicates considerable physical substructure, and $\Sigma_\mathrm{LDR} = 1.44 \pm 0.5$, which shows no significant evidence for mass segregation. These measures both imply that Cyg OB2 is dynamically young as any dynamical interactions between stars will lead to mixing that will erase substructure (causing $Q$ to increase) and promote dynamical mass segregation (causing $\Sigma_\mathrm{LDR}$ to increase). We compare these measures to the predictions of N-body simulations that start from various initial conditions \citep{park14,park16}, measuring these diagnostics at an age of 5~Myr. Regions that start with virial or sub-virial velocity dispersions will quickly collapse down to form compact clusters, erasing their primordial substructure and promoting mass segregation (see Figure~\ref{parameters}) . Even regions that start with super-virial velocity dispersions will undergo some mixing if their initial density is high enough (due to localised correlated velocities) that will increase $\Sigma_\mathrm{LDR}$. To match the observed structural properties of Cyg~OB2 required simulations that start with a low volume density ($< 100$ stars pc$^{-3}$) to prevent close interactions from occurring and increasing $\Sigma_\mathrm{LDR}$. Based on this we concluded that Cyg OB2 formed as a highly substructured, unbound association with a low density, very similar to how we see it now. This work was presented in \citet{wrig14b} and will be extended in the future using the larger and deeper {\it Chandra} Cygnus~OB2 Legacy Survey \citep{wrig14c}.

\subsection{The kinematics of Cyg OB2}

To follow-up on our structural study we sought kinematic measurements to facilitate a dynamical study of Cygnus OB2. The goal of this study was to search for evidence of expansion in the proper motions (PMs), particularly the coherent radial expansion pattern predicted by models of cluster disruption such as residual gas expulsion \citep[e.g.,][]{hill80,lada84,krui11}. Unfortunately at this time the {\it Gaia} satellite had not launched and it was expected to be many years before suitable astrometric data from the satellite would be available. We therefore sought out our own, ground-based astrometry. {\it Gaia} achieves a high PM precision, despite a short baseline, by having an instrument capable of an exceptionally high astrometric precision. While such precision is beyond the capability of most ground-based instruments, this can be compensated for by measuring PMs over a longer baseline. We gathered data from wide field imaging instruments on telescopes around the world spanning a $\sim$15~year baseline and used these to compute PMs that were able to achieve a precision as high as $\sim$0.3~mas/yr.

PMs were obtained for 873 X-ray and spectroscopically selected members of Cyg~OB2 covering the same area as our earlier, structural study. Using a two-component, 2D Gaussian velocity dispersion model we fit the velocity dispersions in the two dimensions as $\sigma_\alpha = 1.89^{+0.07}_{-0.06}$ and $\sigma_\delta = 1.32^{+0.05}_{-0.04}$ mas~yr$^{-1}$, which at a distance of 1.33~kpc \citep{rygl12}, equates to $13.0^{+0.8}_{-0.7}$ and $9.1^{+0.5}_{-0.5}$ km~s$^{-1}$. These values are in approximate agreement with the radial velocity (RV) dispersion measured for the OB stars of $\sigma_{RV} = 8.03 \pm 0.26$ km~s$^{-1}$ \citep{kimi07}, though there is a notable anisotropy present in the 3D velocity dispersions.

\begin{figure}
	\centering
	\includegraphics[width=1\linewidth,angle=270, trim=0 0 0 340,clip]{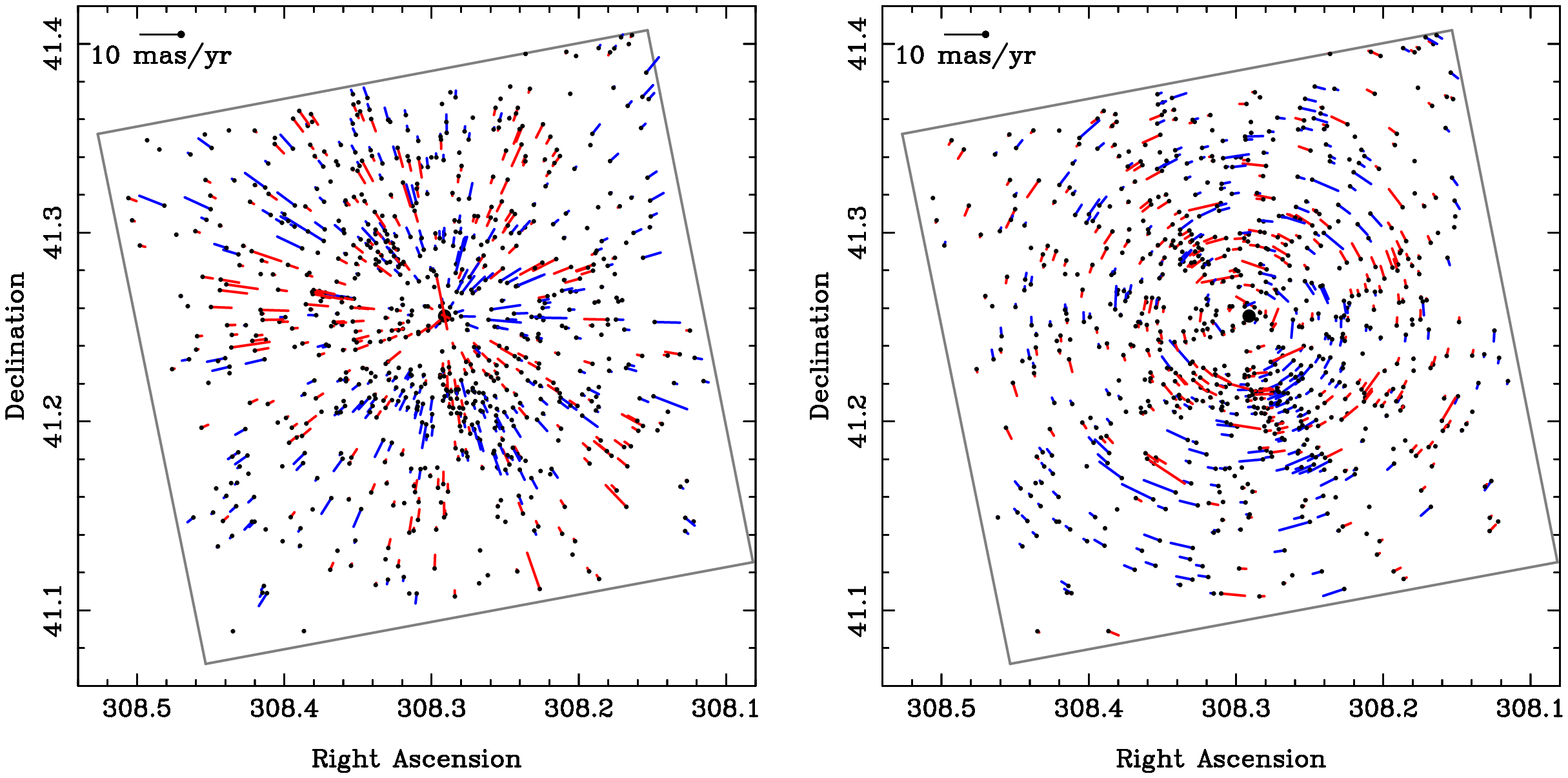}
	\caption{Radial component of the PM vectors for 798 members of Cyg~OB2 (kinematic outliers excluded). The dots show the current positions of the stars while the vectors show their PMs, colour-coded blue if the stars are moving outwards from the centre and red if they are moving inwards. The large black dot shows the nominal association centre. Figure taken from \citet{wrig16}.}
	\label{expansion}
\end{figure}

\begin{figure*}[ht]
	\centering
	\includegraphics[width=450pt]{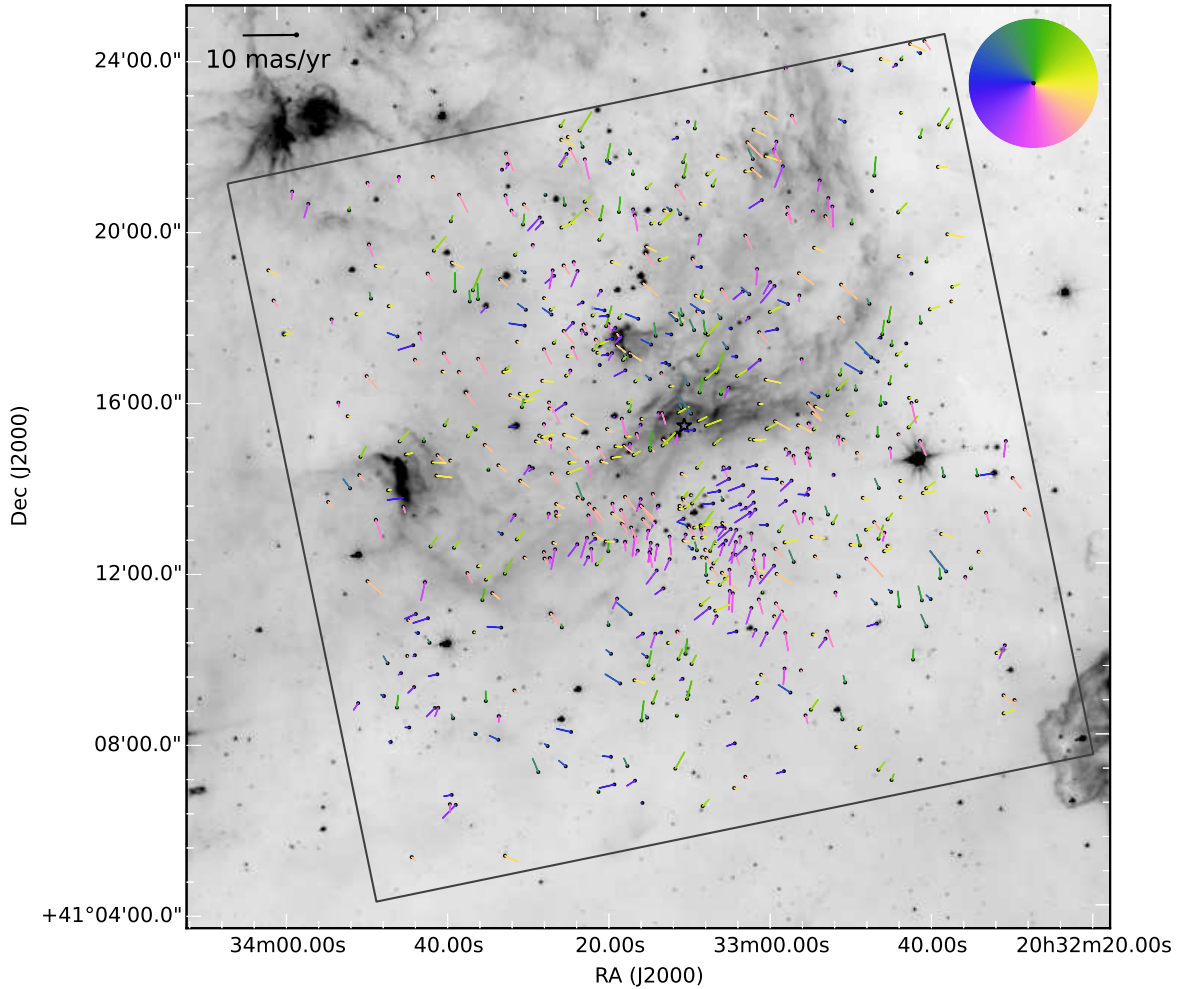}
	\caption{PM vector map for 798 members of Cyg OB2 (kinematic outliers removed). The dots show the current positions of the stars while the vectors show their PMs, colour-coded according to the position angle of their motion (colour wheel shown in top right-hand corner. This highlights the kinematic substructure in the association. A representative 10 mas~yr$^{-1}$ vector is also shown. Figure taken from \citet{wrig16}.}
	\label{substructure}
\end{figure*}

To search for evidence of expansion we divided the PM vectors into radial and transverse components based on the position of each star relative to the nominal centre of the association. For this we used the centre of mass of the sample used here, though the results did not vary when the centre of the association was varied slightly. Approximately $60^{+3}_{-7}$\% of the kinetic energy of the PMs was found in the radial direction, but this is split almost equally between expansion and contraction with $50^{+9}_{-7}$\% of the kinetic energy in the direction of expansion (see Figure~\ref{expansion}). This is in stark contrast to models of residual gas expulsion that suggest that the majority of the kinetic energy should exist in the direction of radial expansion as the cluster explosively expands from a compact origin \citep[e.g.,][]{baum07}.

While the large-scale kinematic structure of the association appears relatively random, with no evidence for cohesive expansion, the small-scale kinematics suggests considerable substructure within the stellar motions. This is shown in Figure~\ref{substructure} where the PMs are coloured according to the position angle of their motion. Stars in the same area of the sky appear to have similar PMs, both in direction and in the amplitude of their vector motion. We use the term {\it kinematic substructure} to describe this observation and suggest that this echoes the physical substructure observed in our previous study. The kinematic substructure is evident on a range of scales, from that of only a few stars up to groups of 10-20 stars or more. Extrapolating our sample to the undetected, lower-mass stars that we assume follow a similar spatial and kinematic distribution suggests these subgroups have masses of 40--400~M$_\odot$. Spatial correlation tests \citep{mora50,gear54} suggest this substructure is real, with significances of 3--10 $\sigma$.

Kinematic substructure such as this is easy to destroy (by dynamical interactions), but difficult to create, and this places strong constraints on the level of dynamical mixing that could have occurred within the association. The velocity dispersions of the stars in these substructures are consistent with them being in virial equilibrium, an observation that is backed up by the fact that the stars are still moving together in these groups despite being $\sim$5~Myr old, which is time enough for an unbound group to disperse. We therefore suggest that these groups represent primordial substructures within the association. This further argues that the association has not undergone a densely clustered phase (during which such substructure would get disrupted or erased) and was likely born with this substructure, potentially in virial equilibrium on small scales, but super-virial on larger scales.

This work was presented in \citet{wrig16}.

\begin{figure*}[h]
	\centering
	\includegraphics[height=450pt,angle=270]{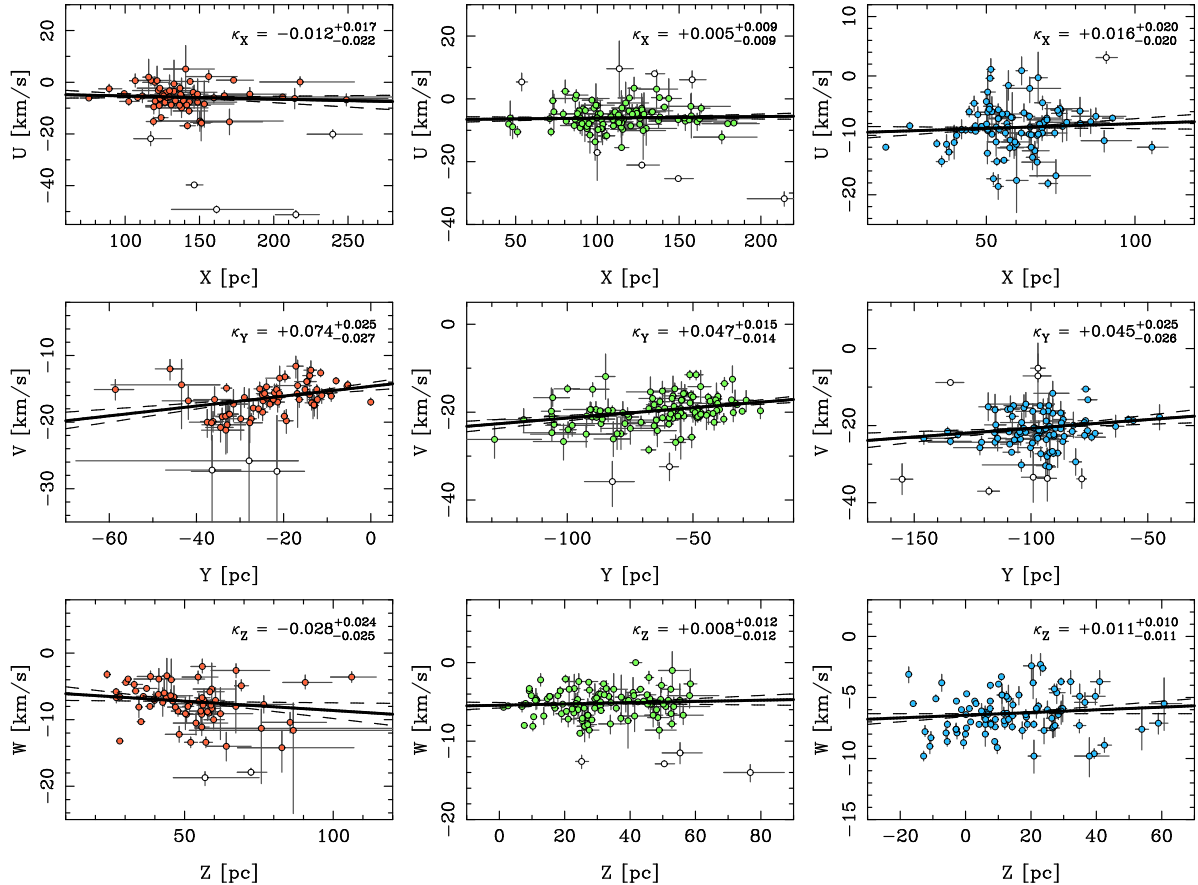}
	\caption{Positions versus velocities along the three Galactic Cartesian axes for each of the three subgroups of Sco-Cen (US on the left, UCL in the centre, and LCC on the right). 1$\sigma$ error bars are shown for all sources. The solid lines show the best-fit linear relationships between the plotted quantities, with 1$\sigma$ uncertainties shown with dashed lines. The best-fitting slopes, $\kappa$, and their uncertainties, are listed in each panel. Figure taken from \citet{wrig18}.}
	\label{3D}
\end{figure*}

\begin{figure*}
	\centering
	\includegraphics[width=500pt]{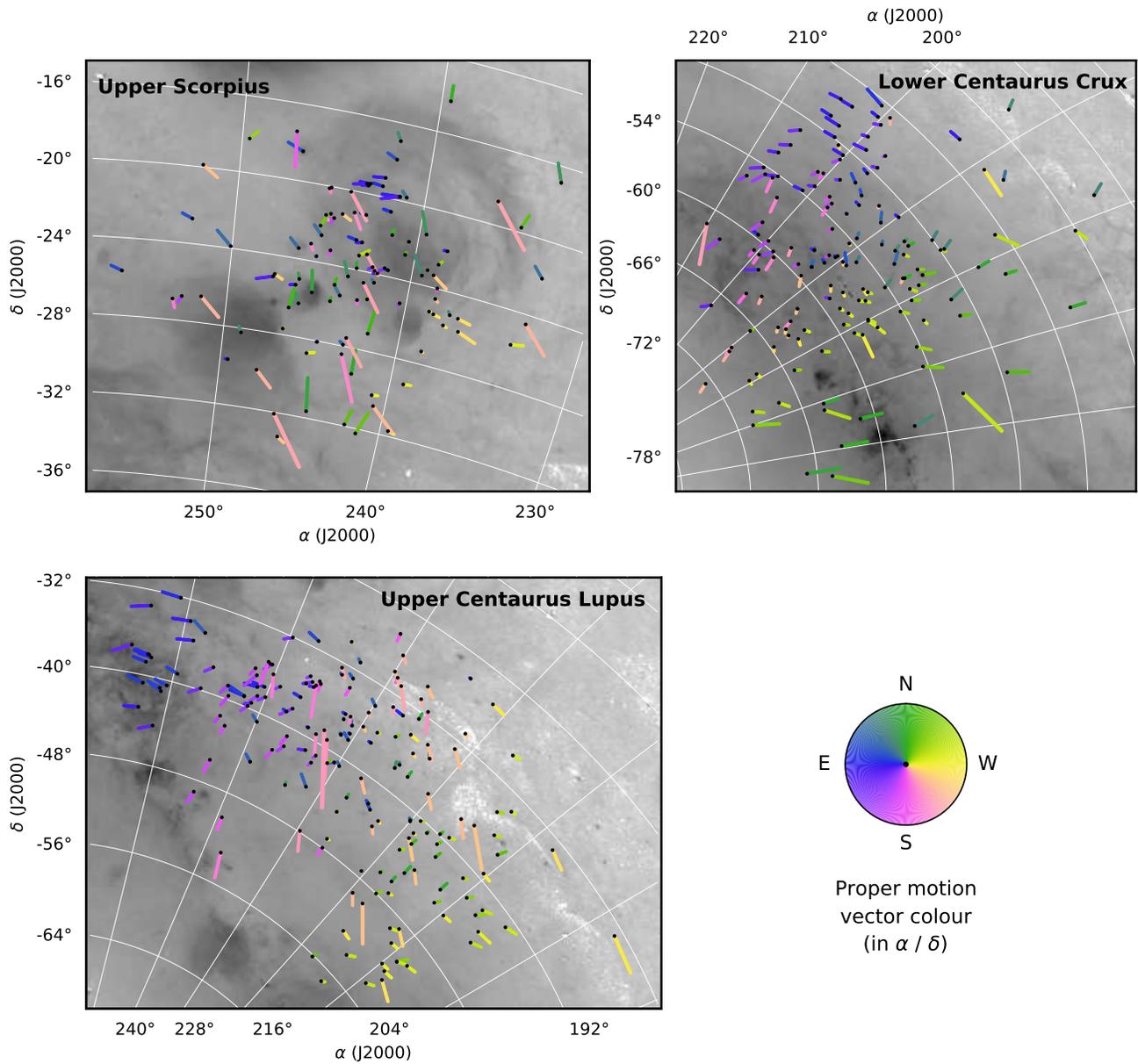}
	\caption{PM vector maps for the three subgroups of Sco-Cen after correction for radial streaming motion using the equations in \citet{brow97}. Points show the current positions of stars and vectors show the proper motions over 0.5~Myrs with the bulk motion of each subgroup subtracted to show the motion in the reference frame of each subgroup. The vectors have been colour-coded based on the position angle of their PM (see colour wheel in lower-right corner) to highlight the kinematic substructure. Figure taken from \citet{wrig18}.}
	\label{corrected_pms}
\end{figure*}

\section{The Scorpius-Centaurus association}

The Scorpius-Centaurus OB association, or Sco-Cen, is the nearest OB association to the Sun \citep[$d \sim 100 - 150$~pc,][]{deze99} and therefore the nearest site of recent, large-scale star formation. The three subgroups, Upper Scorpius (US), Upper Centaurus Lupus (UCL), and Lower Centaurus Crux (LCC), have median ages of 11, 16, and 17~Myr, respectively \citep{peca16}. The entire association is estimated to have a total stellar mass of $\sim$4000~M$_\odot$ \citep{mama02,prei08}, and so while it is considerably less massive than Cyg~OB2 it is nearer and more accessible.

Following the release of {\it Gaia} \citep{prus16} Data Release 1 \citep[DR1,][]{brow16} we gathered PMs for the 433 members of the association in the recent membership list of \citet{rizz11}, which itself is an updated version of the \citet{deze99} membership list from Hipparcos. {\it Gaia} DR1 provides vastly improved PMs for the majority (60\%) of these stars, thanks to PMs calculated from the $\sim$24~yr Hipparcos -- {\it Gaia} baseline. In addition we gathered radial velocities (RVs) from the literature for 63\% of these stars \citep{gont06,khar07,chen11,dahm12}.

\subsection{The expansion of Sco-Cen}

Due to the proximity of Sco-Cen and its large extent on the sky, the expansion of the association cannot be resolved simply by studying the PMs of stars on the plane of the sky. This is both because of the curved surface of the sky (particularly relevant over the almost 90$^\circ$ size of the association on the sky) and the contribution of radial streaming motions to the measured PMs \citep[e.g.,][]{brow97}.

Therefore to assess the evidence for the expansion of the association we adopted numerous other methods, including testing \citet{blaa64}'s linear expansion model, 3D linear expansion tests (Figure~\ref{3D}), comparing the expanding and non-expanding convergent points, tracing back individual stars according to their 3D motions, and performing a correction to the observed PMs for radial streaming motions (Figure~\ref{corrected_pms}). For all of these tests the kinematic data are inconsistent with the three subgroups being the expanded remnants of individual star clusters, with no coherent expansion pattern evident and no evidence that the subgroups had a more compact configuration in the past.

The 3D linear expansion tests shown in Figure~\ref{3D} show velocity plotted against position along each of the three Galactic Cartesian coordinate system axes, $XYZ$. If the subgroups are expanding radially we would expect positive correlations between position and velocity in all three dimensions, but this is not observed. In the $X$ and $Z$ dimensions all three subgroups show either marginally negative or positive slopes that are consistent with a slope of zero. In the $Y$ direction however all three subgroups show significant evidence for expansion at the 2--3 $\sigma$ confidence level. This is not the radial expansion pattern predicted by models such as residual gas expansion. Notably, \citet{mama14} found a similar trend for the $\beta$ Pictoris moving group, which is also expanding preferentially in the $Y$ direction. The surface densities of OB associations and moving groups are low enough that galactic shear may be effective \citep[e.g.,][]{dobb13}, particularly if the shear pattern was imprinted on the molecular gas in the primordial molecular cloud and then inherited by the stars that formed.

Figure~\ref{corrected_pms} shows PM vector maps for the three subgroups of Sco-Cen with the PMs corrected for radial streaming motions using the equations of \citet{brow97}. None of the three subgroups show evidence for a coherent expansion pattern, though there are suggestions of the same kinematic substructure observed in Cyg~OB2. Quantifying the fraction of kinetic energy in the radial part of the proper motions and dividing this between expansion and contraction in each subgroup we find that there is a preference for the motions of stars to be directed away from the centres of each subgroup, with fractions of 59\% (US), 67\% (UCL) and 90\% (LCC) of the kinetic energy in the direction of expansion. This suggests that, while there is not evidence for coherent expansion patterns in the PMs, they do appear to be expanding.

This work was presented in \citet{wrig18}.

\section{Discussion and implications}

The implications from this work are wide-ranging. The idea that all stars form in dense, compact clusters \citep[e.g.,][]{pfal09} is in stark contrast with the observation that OB associations appear to form as highly substructured, super-virial, and generally low-density agglomerates of young stars. This also means that the very massive stars that are common in OB associations, for example in Cyg~OB2 \citep[with masses up to $\sim$100~M$_\odot$,][]{wrig15a} must have formed in either a relatively small cluster or a relatively low-density environment. This is inconsistent with the idea that massive stars must form in dense star clusters \citep[e.g.,][]{bonn01,york02,zinn07} or that there is an important physical correlation between the cluster mass and the mass of the most massive star in the cluster \citep{weid06}.

The lack of evidence for the radial expansion pattern predicted by theories of cluster disruption such as residual gas expulsion \citep{hill80,baum07} is also at odds with these observations, suggesting that such mechanisms either do not result in such a kinematic pattern or that those mechanisms are not as effective in disrupting star clusters as once thought. The fact that the kinematic substructures identified in Cyg~OB2 are close to virial equilibrium also suggests that residual gas expulsion has not had a significant impact on the dynamics of small-scale systems such as those.

Finally, if OB associations originate as relatively low density systems then this has implications for the processing of binary and planetary systems that form in those associations. For example, observations of Cyg~OB2 show an excess of wide binaries amongst the massive star population in the association (Caballero-Nieves et al. {\it in prep.}), which \citet{grif18} argue supports the picture put forward by \citet{wrig14b} that Cyg~OB2 did not form as a single dense cluster but as a lower-density, substructured association. The low density conditions in such OB associations could therefore mean less disruption or processing of wide binaries and planetary systems relative to those formed in dense clusters. The proximity of the Sco-Cen association will allow this to be explored in more detail in the future.

\section{Summary}

OB associations have long been thought to be the expanded remnants of dense and compact star clusters. We find, from both structural and kinematic studies of multiple OB associations, that while such systems are expanding they are not expanding from compact initial conditions, but from extended and substructured distributions. This has considerable and wide-ranging implications, from the star formation process to the formation of planetary systems.

\section*{Acknowledgments}
{I would like to thank the organisers of all Cool Stars conferences, past and present, for their hard work organising this excellent series of conferences, and for giving me the opportunity to speak. I would also like to thank David Barrado, Emmanuel Bertin, Herve Bouy, Jean-Charles Cuillandre, Janet Drew, Jeremy Drake, Simon Goodwin, Rob Jeffries, Eric Mamajek, Richard Parker and Luis Manuel Sarro for their contributions to this work.}

\bibliographystyle{cs20proc}
\bibliography{bibliography}

\end{document}